\documentclass[10pt,conference]{IEEEtran}
\usepackage[T1]{fontenc}
\usepackage[utf8]{inputenc}
\usepackage[USenglish]{babel}
\usepackage{url}
\usepackage{breakurl} 
\usepackage[bookmarks=false, bookmarksnumbered=true, hypertexnames=false, breaklinks=true, linkbordercolor={1 1 1}, pdfborder={0 0 0}]{hyperref}

\let\labelindent\relax
\usepackage{cleveref}
\crefformat{footnote}{#2\footnotemark[#1]#3}
\usepackage{enumitem}
\usepackage{microtype}
\usepackage{flushend}
\newcommand{\rqthree}{What are the experienced consequences of unhappiness among software developers while developing software?}

\begin{document}
%
\title{Consequences of Unhappiness While Developing Software}

\author{
    \IEEEauthorblockN{
        Daniel Graziotin\IEEEauthorrefmark{1},
        Fabian Fagerholm\IEEEauthorrefmark{2}, 
        Xiaofeng Wang\IEEEauthorrefmark{3} and
        Pekka Abrahamsson\IEEEauthorrefmark{4}
    }
    \IEEEauthorblockA{
       \IEEEauthorrefmark{1}Institute of Software Technology, University of Stuttgart, Germany, \href{mailto:daniel.graziotin@informatik.uni-stuttgart.de}{daniel.graziotin@informatik.uni-stuttgart.de}
    }
    \IEEEauthorblockA{
       \IEEEauthorrefmark{2}Department of Computer Science, University of Helsinki, Finland, \href{mailto:fabian.fagerholm@helsinki.fi}{fabian.fagerholm@helsinki.fi}
    }
    \IEEEauthorblockA{
       \IEEEauthorrefmark{3}Faculty of Computer Science, Free University of Bozen-Bolzano, Italy, \href{mailto:xiaofeng.wang@unibz.it}{xiaofeng.wang@unibz.it}
    }
    \IEEEauthorblockA{
       \IEEEauthorrefmark{4}Department of Computer and Information Science (IDI), NTNU, Trondheim, Norway, \href{mailto:pekkaa@ntnu.no}{pekkaa@ntnu.no}
    }
}

\maketitle

\begin{abstract}
The growing literature on affect among software developers mostly reports on the linkage between happiness, software quality, and developer productivity. Understanding the positive side of happiness -- positive emotions and moods -- is an attractive and important endeavor. Scholars in industrial and organizational psychology have suggested that also studying the negative side -- unhappiness -- could lead to cost-effective ways of enhancing working conditions, job performance, and to limiting the occurrence of psychological disorders. Our comprehension of the consequences of (un)happiness among developers is still too shallow, and is mainly expressed in terms of development productivity and software quality. In this paper, we attempt to uncover the experienced consequences of unhappiness among software developers. Using qualitative data analysis of the responses given by 181 questionnaire participants, we identified 49 consequences of unhappiness while doing software development. We found detrimental consequences on developers' mental well-being, the software development process, and the produced artifacts. Our classification scheme, available as open data, will spawn new happiness research opportunities of cause-effect type, and it can act as a guideline for practitioners for identifying damaging effects of unhappiness and for fostering happiness on the job.
\end{abstract}


%
\IEEEpeerreviewmaketitle

\section{Introduction}

Software companies often gladly promote the idea of flourishing happiness among developers, knowingly or accidentally attempting to enact the happy-productive worker thesis~\cite{Zelenski2008}. The happiness of all stakeholders involved in software development is an essential element of company success~\cite{Denning2012b}. Recent research within the scope of behavioral software engineering~\cite{Lenberg:2015bj} has highlighted the relationship between software developer happiness and work-related constructs such as performance and productivity~\cite{graziotin2015you,Graziotin2014PEERJ,graziotin2015feelings,Fagerholm2015,Muller2015,Ortu2015a}, quality~\cite{Khan2010,Destefanis2016}, and the social interactions between developers~\cite{Novielli:2015km}. Most of the studies to date have investigated the positive side of happiness.

While happiness, for the individual, is inherently subjective, research shows that it can be studied objectively. Objective happiness can be construed as the difference between experienced positive affect and experienced negative affect~\cite{Diener:1999cl, Kahneman:1999ck}. Thus, maximizing happiness may be achieved by either maximizing positive experiences or minimizing negative experiences (or both). Focusing on the negative may already be intuitive to many developers. 

It is a common occurrence that developers share horror stories about their working experience~\cite{Graziotin2014IEEESW}. Managers in the software profession would benefit from greater understanding of the nature and dynamics of unhappiness among developers, and they could take action to prevent dysfunctional responses among employees~\cite{Vecchio:2000ei}. Further understanding of the benefits of limiting negative experiences on the job in general has been called for~\cite{Diener:1999cl}. 

We aim to broaden the understanding of the unhappiness of software developers, and are conducting a series of studies using a large-scale quantitative and qualitative survey of software developers. In those studies, we assess the happiness in the developer population, the causes of (un)happiness, and what the consequences of those experiences are. The study we are describing in the present paper is the first of the series\footnote{We are offering a preview of the results of the present study in an ICSE'17 poster as well~\cite{Graziotin:2017}.}. 

In the present article, we focus on the consequences of unhappiness, and investigate the following research question which is based on the existing literature:

\vspace{1mm}

\noindent RQ: \emph{\rqthree}

\vspace{1mm}

We report 49 consequences of unhappiness that we identified. The consequences concern developers themselves in the form of cognitive and behavioral changes, and external outcomes related to the software development process and artifacts.

\section{Related Work}
\label{sec:related:background_and_related_work}

From a \emph{hedonistic} viewpoint, \textbf{happiness} is a sequence of experiential episodes \cite{Haybron2001} and being happy (unhappy) is associated with frequent experiences of positive (negative) affect~\cite{Diener2010}.\footnote{Alternative views of happiness exist, e.g. Aristoteles' \emph{eudaimonia}: a person is happy because (s)he conducts a satisfactory life full of quality \cite{Haybron2005}. A review of affect theories is given in \cite{Graziotin2015SSE} and the role of the centrality of affect and happiness is discussed in \cite{graziotin2015you}.} A considerable increase in the interest of studying affect and happiness among software developers is visible over the last five years, although the research is in its infancy; many theoretical and methodological issues remain in software engineering research, as illustrated by Graziotin et al.~\cite{Graziotin2015SSE,graziotin2015affect} and by Novielli et al.~\cite{Novielli:2015km}.

Several studies have attempted to elucidate the complex relationship between happiness (and more generally, affect) and performance in the context of software development. In a study of the affect associated with eliciting requirements, investigating 65 user requirements from two projects, high activation and low pleasure levels were shown to be predictors of high versioning requirements~\cite{ColomoPalacios:2011jq}. Pleasure increased over time with each new version, while activation decreased. 

Theory-building is an important part of software engineering research, and theories regarding affect can inform further empirical studies. One such theory is an explanatory process theory of the impact of affect on development performance~\cite{graziotin2015you}. The theory was formed by qualitative analysis of interview data, communications, and observations of two software developers in the same project. The concept of attractors -- affective experiences that earn importance and priority to a developer's cognitive system -- was theorized to have the biggest impact on development performance.

Correlational experiments have found a positive relationship between happiness and positive emotions arising from a development task~\cite{Graziotin2014PEERJ,graziotin2015feelings}, and between problem-solving performance and development task productivity~\cite{Muller2015}. Affect has also been shown to impact debugging performance: in a controlled experiment where participants were asked to write a trace of algorithm execution, induced high pleasure and activation affect were found to be associated with high debugging performance~\cite{Khan2010}.

Further evidence for the link that developers experience between emotion and performance is provided in a survey with 49 developers, assessing emotions they perceived to influence their own productivity~\cite{Wrobel2013}. Positive affective states were perceived to be those that enhance development productivity. The negative affect most prevalently perceived was frustration, which was also the one perceived to deteriorate productivity the most.

In all studies investigating the happiness of developers in different forms and its impact on performance, the findings point to a positive relationship. The findings are similar when it comes to software quality. A series of studies using software repository mining found links between affect, emotions, and politeness, and software quality~\cite{Ortu2015a, Destefanis2016, Mantyla2016a}. Happiness in terms of frequent positive affect and positive emotions was found to be associated with shorter issue fixing time~\cite{Ortu2015a}. The level of arousal, which, when high, is associated with anxiety and burnout, was found the be associated with issue priority~\cite{Mantyla2016a}. Politeness in requests for resolving issues was correlated with lower resolution time~\cite{Destefanis2016}.

\section{Method}
\label{sec:method:method}

The present study is the first in a series of inquiries that we conducted on data from a large-scale survey of developers. The overall research project employs a mixed research method, with elements of both quantitative and qualitative research~\cite{Creswell2009}. The present study, however, is purely qualitative.

\subsubsection{Sampling Strategy}
\label{sec:method:data-retrieval}

We used GitHub as an avenue for reaching software developers that would represent the population of developers \textit{well enough}, following several previous studies (e.g.~\cite{Gousios:2016hj}). GitHub has more than 30 million visitors each month~\cite{Doll:2015uv} and is, as far as we can tell, the largest social coding community in the world. Software developers using GitHub work on a wide variety of projects, ranging from open source to proprietary software and from solo work to work done in companies and communities.

We extracted a set of developer contacts from the \emph{\href{https://githubarchive.org}{GitHub Archive}}, which stores public events occurring in GitHub. We retrieved event data for a period of six months, and extracted email addresses, given names, company names, developer locations, and the repository name associated with each event.

\subsubsection{Survey Design}

We designed a survey consisting of (1) questions regarding demographics, (2) one question with the Scale of Positive and Negative Experience (SPANE~\cite{Diener2010}) with 12 items assessing happiness, and (3) two open-ended questions asking for experienced causes and consequences of positive and negative affect when developing software. The questionnaire is described in an online appendix~\cite{Graziotin:2016ck}. We piloted the questionnaire thrice, allowing us to estimate and improve response rates by refining the questions and invitation email. No data from the pilots were retained in the final data set and pilot participants did not participate in the final round. The present article covers the results related to the open-ended questions regarding the consequences of negative affect (thus, unhappiness) while developing software.

\subsubsection{Analysis}

We qualitatively analyzed the cleaned data for the open-ended questions. In order to answer the RQ of this paper, we used the responses for the question on consequences. We developed a coding strategy, applying open coding, axial coding, and selective coding as defined by Corbin and Strauss' Grounded Theory~\cite{Corbin2008} as follows\footnote{For a review and guidelines of Grounded Theory in software engineering research, see \cite{Stol:2016:GTS:2884781.2884833}}. The first three authors each coded the same set of 50 responses using a line-by-line strategy. We then compared the coding structure and strategy and reached an agreement, i.e., a shared axial coding scheme. We took the individual developer as the starting point and unit of observation and analysis, and based the construction of theoretical categories on a model~\cite{Curtis:1988:FSS:50087.50089} of constructs that are \emph{internal} or \emph{external} to the developer. The internal category concerns the developer's own being, while the external category contains artifacts, processes, and people as subcategories. We then divided the data evenly among the first three researchers and proceeded to open code them. We monitored our progress and further discussed the coding scheme and strategy in weekly meetings. We finally merged the codes and conducted a last round of selective coding. An example illustrating the coding process is available online~\cite{Graziotin:2016ck}. All analysis was done using \textit{NVIVO 11}.

\section{Results}
\label{sec:results:results}

In this section, we summarize the results of our investigation. We first show descriptive statistics describing the demographics of the participants. We then proceed to the qualitative data related to our RQ.

\subsection{Descriptive Statistics}
\label{ssec:results:descriptive}
We obtained $181$ valid and complete---that is, after data cleaning---responses related to our RQ, which resulted in $172$ male participants (95\%) and 8 female (4\%). The remaining participant declared their gender as \textit{other / prefer not to disclose}. The mean for the year of birth was 1984 (standard deviation (sd)=8.27), while the median was 1986. A wide range of nationalities was represented, with 45 countries. 141 (78\%) participants were professional software developers, 7\% were students, and 13\% were in other roles (such as manager, CEO, CTO, and academic researcher). The remaining participants were non-employed and not students. The participants declared a mean of 8.22 years (sd=7.83) software development working experience, with a median of 5 years.

\subsection*{What are the Experienced Consequences of Unhappiness Among Software Developers While Developing Software?}
\label{ssec:resultsrq3}

We now provide a summary of the elicited consequences of unhappiness while developing software. 
We identified 254 codes related to the consequences of unhappiness, which resulted in 49 consequences, divided into 16 categories and sub-categories. Because of space limitations, we report here the most frequent codes. The entire dataset is available as archived open data \cite{Graziotin:2016ck}. We found support for Curtis et al. \cite{Curtis:1988:FSS:50087.50089} internal category, which we label \emph{developer's own being} (112 references), and the external categories \emph{process} (106) and \emph{artifact} (36).

\subsection{Internal Consequences---Developer's Own Being}
\label{sssec:resultsrq3:individual_consequences}
The developer's own being-related factors do not demonstrate a clear structure. This to some extent reflects the versatile states of mind of developers and the feelings they could have while they develop software.

The most significant consequences of unhappiness for the \textit{developers' own being} are: low cognitive performance, mental unease or disorder, and low motivation.

\textbf{Low cognitive performance} is a category to group all those consequences related to low mental performance, such as low focus: ``\textit{[\ldots] the negative feelings lead to not thinking things through as clearly as I would have if the feeling of frustration was not present}''; cognitive skills dropping off: ``\textit{My software dev skills dropped off as I became more and more frustrated until I eventually closed it off and came back the next day to work on it}''; and general mental fatigue: ``\textit{Getting frustrated and sloppy}''.

The \textbf{mental unease or disorder} category collects all those consequences that threaten mental health\footnote{\label{trained-psych}In this study, we report what the participants stated, but we remind readers that only trained psychologists and psychiatrists should treat or diagnose mental disorders.}. The participants reported that unhappiness while developing software is a cause of, in order of frequency, stress and burnout: ``\textit{[\ldots] only reason of my failure due of burnout}''; anxiety: ``\textit{These kinds of situations make me feel panicky}''; low self-esteem: ``\textit{If I feel particularly lost on a certain task, I may sometimes begin to question my overall ability to be a good programmer}''; and sadness. Participants mentioned depression as in feeling depressed, e.g., ``\textit{feels like a black fog of depression surrounds you and the project}'' or ``\textit{I get depressed}''.

\textbf{Low motivation} is also an important consequence of unhappiness for software developers. Motivation is a set of psychological processes that cause the mental activation, direction, and persistence of voluntary actions that are goal directed \cite{Mitchell1982}. Motivation has been the subject of study in software engineering literature (e.g, \cite{Franca2014b}), and we reported that affective experiences are related to motivation even though they are not the same construct \cite{graziotin2015affect}. The participants were clear in stating that unhappiness leads to low motivation for developing software, e.g, ``\textit{[the unhappiness] has left me feeling very stupid and as a result I have no leadership skills, no desire to participate and feel like I'm being forced to code to live as a kind of punishment. [\ldots]}'', or ``\textit{Also, I'm working at a really slow pace [\ldots] because I'm just not as engaged with the work}''.

\textbf{Work withdrawal} is a very destructive consequence of unhappiness, and it emerged often among the responses. Work withdrawal is a family of behaviors that is defined as employees’ attempts to  remove themselves, either temporarily or permanently, from quotidian work tasks \cite{Miner2010}. The gravity of this consequence ranged from switching to another task, e.g, ``\textit{[\ldots] you spend like 2 hours investigating on Google for a similar issue and how it was resolved, you find nothing, desperation kicks in. It clouds your mind and need to do other things to clear it}'', to considering quitting developing software, ``\textit{I really start to doubt myself and question whether I'm fit to be a software developer in the first place}'', or even, ``\textit{I left the company}''.

\subsection{External Consequences---Process }

The category of \textit{process} collects those unhappiness consequences that are related to a software development process, endeavor, or set of practices that is not explicitly tied up to an artifact (see Section \ref{sssec:resultsrq3:artifactoriented_consequences}).

\textbf{Low productivity} is a category for grouping all consequences of unhappiness related to performance and productivity losses\footnote{See \cite{graziotin2015you} and \cite{Fagerholm2015} for our stance on a definition of productivity and performance in software engineering.}. The codes within this category were ranging from very simple and clear ``\textit{productivity drops}'', ``\textit{[negative experience] definitely makes me work slower}'' to more articulated ``\textit{[unhappiness] made it harder or impossible to come up with solutions or with good solutions}'', ``\textit{[\ldots], and [the negative experience] slowed my progress because of the negative feeling toward the feature}''.

Unhappiness was reported to be causing \textbf{delay} in executing process activities: ``\textit{In both cases [negative experiences] the emotional toll on me caused delays to the project}''. Unhappiness causes glitches to communication activities and a disorganized process: ``\textit{Miscommunication and disorganization made it very difficult to meet deadlines}''.

Developers declared that unhappiness caused them to \textbf{deviate from the process} or the agreed set of practices. Specifically, unhappiness makes developers compromise in terms of actions, in order to just get rid of the job: ``\textit{In these instances my development tended towards immediate and quick 'ugly' solutions}''.
Developers see their quality of the code compromised (Section \ref{sssec:resultsrq3:artifactoriented_consequences}) but also decide to take shortcuts when enacting a software process, compromising the quality of the process itself: ``\textit{[\ldots] can lead to working long hours and trying to find shortcuts. I'm sure this does not lead to the best solution, just a quick one}''. The process adherence can suffer due to communication aspects, too: ``\textit{my development was influenced by [negative affect] in that it caused me to tighten up communications and attempt to force resolution of the difficulties}''.

Somehow related to the process deviation is the \textbf{broken flow} category. Unhappiness causes developers to interrupt the flow, as described by Csikszentmihalyi \cite{Csikszentmihalyi1997} and investigated by Müller and Fritz \cite{Muller2015} as an attention state of progressing and concentration. As put by a participant, `\textit{things like that [of unhappiness] often cause long delays, or cause one getting out of the flow, making it difficult to pick up the work again where one has left off. }''. Unhappiness and the broken flow make developers stand up and ``\textit{[\ldots] make me quit and take a break}''; the feeling of getting stuck is constant.

\subsection{External Consequences---Artifact-oriented}
\label{sssec:resultsrq3:artifactoriented_consequences}
The category of \textit{artifact-oriented} consequences groups all those consequences that are directly related to a development product, e.g., software code, requirements, and to working with it. As expected by the foci of previous research, the most important consequence of unhappiness of software developers was low software quality.

\textbf{Low code quality} represents the consequences of unhappiness of developers that are related to deterioration of the artifacts' quality. The participants reported that ``\textit{eventually [due to negative experiences], code quality cannot be assured. So this will make my code messy and more bug can be found in it}''. but also mentioned making the code less performant, or ``\textit{As a result my code becomes sloppier}''. Moreover, participants also felt that they could discharge quality practices, e.g, ``\textit{so I cannot follow the standard design pattern}'', as a way to cope with the negative experiences.

\textbf{Discharging code} could be seen as an extreme case of productivity and quality drop. We found some instances of participants who destroyed the task-related codebase, e.g, ``\textit{I deleted the code that I was writing because I was a bit angry}'', up to deleting entire projects: ``\textit{I have deleted entire projects to start over with code that didn't seem to be going in a wrong direction}''. 

\section{Discussion}\label{sec:discussion:discussion}

In the quest for answering our RQ, we have shown that the unhappiness of developers negatively impacts several important software engineering outcomes. Productivity and performance are the aspects which suffer most from unhappy developers. When grouping low cognitive performance and process-related productivity codes, about 40\% of the related in-text references deal with productivity and performance drops. Those results are in line with and support the related work in software engineering research \cite{graziotin2015you, Graziotin2014PEERJ, graziotin2015feelings, Muller2015, Khan2010, Wrobel2013} which quantified the relationship or attempted to explain the link. 

Our results show that unhappiness while programming may be source of several mental-related issues that are known to be of detrimental effect to the work environment. We found situations of mental unease, e.g., low self-esteem, high anxiety, burnout, and stress. Initial software engineering research on the latter two has started (e.g., \cite{Mantyla2016a}), but the related work in psychology is comprehensive and alarming in regards to how disruptive these issues are on well-being. Furthermore, our data has also shown mentions of possible mental disorders such as depression\cref{trained-psych}.

Continuing with issues related to intellectual performance, unhappiness appears to also bring down motivation among developers, which is a critical force in software engineering activities \cite{Franca2014b}. Negative experiences and negative affect are also experienced to be causes of work withdrawal. Psychology research has recently started to investigate affect versus work withdrawal (e.g., \cite{Miner2010}), but we are not aware of related software engineering research. Our data has shown that work withdrawal causes developers to distance themselves from the task that raises unhappiness up to the point of quitting jobs.

Finally, unhappiness makes developers take process-related shortcuts (i.e., to ``cut corners''). These deviations are often mentioned to cause issues in terms of software quality. While a few studies have been conducted on the affect of developers and its impact on software quality (e.g, \cite{Khan2010,Destefanis2016,Ortu:2016gz,Ortu2015a}), we encourage future research on the matter.

\subsection{Limitations}

We elicited the causes of unhappiness of software developers using qualitative data analysis techniques. Whether causality can be inferred from other than controlled experiments, e.g., eliciting experiences from introspection in the context of qualitative research, is a matter of debate \cite{Creswell2009,Djamba:2002jo,Glaser:2013ha}. However, several authors, e.g., \cite{Glaser:2013ha}, take the stance that qualitative data analysis is able to infer causality from experience of \emph{human} participants, provided that there is a strong methodology for data gathering and analysis. In our case, we paid careful attention to Grounded Theory coding methodology \cite{Corbin2008} in order to strengthen the validity of our results. Furthermore, eliciting the consequences of unhappiness as experienced by developers themselves was our research goal. As the consequences come from first-hand reports, we argue that they accurately represent the respondents' views. Our study does not show any general relationship between any specific consequences; only experienced consequences of unhappiness are claimed.

Our sample of software developers using GitHub is limited in size and with respect to representativeness of developers at large. Our dataset (see Section \ref{sec:method:data-retrieval}) contains accounts with public activity during a six-month period. This may result in a bias as developers who prefer not to display their work in public would not be present in the data set; nor would developers whose work is done in companies' internal systems. The six-month time period, however, is less of an issue as very inactive developers are of less interest to our study -- but they may differ in terms of how they view consequences of unhappiness. Replication using different data sources and collection methods is needed to validate our results in these scenarios.

It might be the case that the ``GitHub population of developers'' is slightly younger than developers in general, but to our knowledge no supporting empirical evidence exists. GitHub is a reliable source for obtaining software engineering research data, as it allows replication of this study on the same or different populations. The GitHub community is large (30 million visitors per month~\cite{Doll:2015uv}) and diverse in terms terms of team size, type of software, and several other characteristics. Our sample is similarly diverse and is balanced in terms of demographic characteristics, including participant role, age, experience, work type, company size, and students versus workers. One exception is gender: our sample is strongly unbalanced towards males. We believe, alas, that our sample is representative in terms of gender as well, as it is a known problem that software engineering roles are predominantly filled by males \cite{Ortu:2016gz,Terrell:2016dq,Ford:2016}, although recent research is attempting to tackle the issue.

\subsection{Recommendations for Practitioners}

We believe that our discovered consequences of unhappiness of developers should be of interest to practitioners working as managers, team leaders, but also team members. While the present paper contains the most prominent factors, we made the entire list available as archived open data \cite{Graziotin:2016ck}. Practitioners in leadership positions should attempt to foster happiness of software development teams by limiting their unhappiness. The benefits of fostering happiness among developers were empirically demonstrated in past research, and they especially highlight software development productivity and software quality boosts. With our results, we add that addressing unhappiness will limit the damage in terms of several factors at the individual, artifact, and process level. Moreover, previous research \cite{graziotin2015you} has suggested that intervening on the affect of developers might have relatively low costs and astonishing benefits.

\subsection{Implications for Researchers}

We believe that the results of the present work could be adopted as the basis of several research directions. Our study has the potential to open up new avenues in software engineering research based on the discovered factors (e.g., work withdrawal and affect of developers). Also, all the factors we have reported are the end part of an experienced causality chain with unhappiness as the antecedent. Future studies should attempt to seek a quantification of the chain.

\section{Conclusion}In this paper, we presented the results of an analysis of the experienced consequences of unhappiness among software developers while developing software. The complete results are archived and available as open data \cite{Graziotin:2016ck}. The consequences are grouped into the main categories of internal -- developer's own being -- and external -- process and artifact. The highest impact is experienced to be on programming productivity as expressed by cognitive performance, including creativity and flow, and process-related performance. We found several instances of job-related adverse effects and even indications of mental disorders: work withdrawal, stress, anxiety, burnout, and depression\cref{trained-psych}. 

Our recommendation to practitioners, including managers and team leaders, is to utilize our list of consequences and the explanations offered by the present paper to start their quest for enhancing the working conditions of software developers. The consequences, in particular, offer interesting angles which managers should reflect on and look out for in their workforce.

We believe that our study results are of immediate application in future academic work. The results set theoretical foundations for causality studies and inspiration for novel research activities in software engineering.

The present study enforces the stance that many aspects of software engineering research require approaches from the behavioral and social sciences; we believe there is a need in future academic discussions to reflect on how software engineering research can be characterized in such terms. Developers are prone to share work-related horror stories on a daily basis, and we believe that their job conditions are often overlooked. With our past and present research activities, we hope we can contribute towards a higher well-being of software engineers, while enhancing the amount and quality of their job outputs.

\section*{Acknowledgment}
The authors would like to thank all those who participated in this study. 
Daniel Graziotin has been supported by the Alexander von Humboldt (AvH) Foundation.



%
\bibliographystyle{IEEEtran}
\bibliography{references}

\begin{thebibliography}{10}
\providecommand{\url}[1]{#1}
\csname url@samestyle\endcsname
\providecommand{\newblock}{\relax}
\providecommand{\bibinfo}[2]{#2}
\providecommand{\BIBentrySTDinterwordspacing}{\spaceskip=0pt\relax}
\providecommand{\BIBentryALTinterwordstretchfactor}{4}
\providecommand{\BIBentryALTinterwordspacing}{\spaceskip=\fontdimen2\font plus
\BIBentryALTinterwordstretchfactor\fontdimen3\font minus
  \fontdimen4\font\relax}
\providecommand{\BIBforeignlanguage}[2]{{%
\expandafter\ifx\csname l@#1\endcsname\relax
\typeout{** WARNING: IEEEtran.bst: No hyphenation pattern has been}%
\typeout{** loaded for the language `#1'. Using the pattern for}%
\typeout{** the default language instead.}%
\else
\language=\csname l@#1\endcsname
\fi
#2}}
\providecommand{\BIBdecl}{\relax}
\BIBdecl

\bibitem{Zelenski2008}
J.~M. Zelenski, S.~A. Murphy, and D.~A. Jenkins, ``{The Happy-Productive Worker
  Thesis Revisited},'' \emph{Journal of Happiness Studies}, vol.~9, no.~4, pp.
  521--537, 2008.

\bibitem{Denning2012b}
P.~J. Denning, ``{Moods},'' \emph{Communications of the ACM}, vol.~55, no.~12,
  p.~33, Dec. 2012.

\bibitem{Lenberg:2015bj}
{Lenberg, Per}, {Feldt, Robert}, and {Wallgren, Lars-G{\"o}ran}, ``{Behavioral
  software engineering},'' \emph{Journal of Systems and Software}, vol. 107,
  no.~C, pp. 15--37, Sep. 2015.

\bibitem{graziotin2015you}
D.~Graziotin, X.~Wang, and P.~Abrahamsson, ``{How do you feel, developer? An
  explanatory theory of the impact of affects on programming performance},''
  \emph{PeerJ Computer Science}, vol.~1, no.~1, p. e18, Aug. 2015.

\bibitem{Graziotin2014PEERJ}
------, ``{Happy software developers solve problems better: psychological
  measurements in empirical software engineering},'' \emph{PeerJ}, vol.~2, p.
  e289, 2014.

\bibitem{graziotin2015feelings}
------, ``{Do feelings matter? On the correlation of affects and the
  self-assessed productivity in software engineering},'' \emph{Journal of
  Software: Evolution and Process}, vol.~27, no.~7, pp. 467--487, 2014.

\bibitem{Fagerholm2015}
F.~Fagerholm, M.~Ikonen, P.~Kettunen, J.~M{\"u}nch, V.~Roto, and
  P.~Abrahamsson, ``{Performance Alignment Work: How software developers
  experience the continuous adaptation of team performance in Lean and Agile
  environments},'' \emph{Information and Software Technology}, vol.~64, pp.
  132--147, 2015.

\bibitem{Muller2015}
S.~C. Muller and T.~Fritz, ``{Stuck and Frustrated or in Flow and Happy:
  Sensing Developers' Emotions and Progress},'' in \emph{Proc.\ 37th IEEE
  International Conference on Software Engineering}.\hskip 1em plus 0.5em minus
  0.4em\relax IEEE, May 2015, pp. 688--699.

\bibitem{Ortu2015a}
M.~Ortu, B.~Adams, G.~Destefanis, P.~Tourani, M.~Marchesi, and R.~Tonelli,
  ``{Are Bullies More Productive? Empirical Study of Affectiveness vs. Issue
  Fixing Time},'' in \emph{2015 IEEE/ACM 12th Working Conference on Mining
  Software Repositories (MSR)}.\hskip 1em plus 0.5em minus 0.4em\relax IEEE,
  2015, pp. 303--313.

\bibitem{Khan2010}
I.~A. Khan, W.-P. Brinkman, and R.~M. Hierons, ``{Do moods affect
  programmers{\textquoteright} debug performance?}'' \emph{Cognition,
  Technology {\&} Work}, vol.~13, no.~4, pp. 245--258, 2010.

\bibitem{Destefanis2016}
G.~Destefanis, M.~Ortu, S.~Counsell, S.~Swift, M.~Marchesi, and R.~Tonelli,
  ``{Software development: do good manners matter?}'' \emph{PeerJ Computer
  Science}, vol.~2, no.~4, p. e73, 2016.

\bibitem{Novielli:2015km}
N.~Novielli, F.~Calefato, and F.~Lanubile, ``{The challenges of sentiment
  detection in the social programmer ecosystem},'' in \emph{SSE 2015
  Proceedings of the 7th International Workshop on Social Software
  Engineering}.\hskip 1em plus 0.5em minus 0.4em\relax New York, New York, USA:
  ACM, Sep. 2015.

\bibitem{Diener:1999cl}
E.~Diener, E.~M. Suh, R.~E. Lucas, and H.~L. Smith, ``{Subjective well-being:
  Three decades of progress.}'' \emph{Psychological Bulletin}, vol. 125, no.~2,
  pp. 276--302, 1999.

\bibitem{Kahneman:1999ck}
D.~Kahneman, ``{Objective Happiness},'' in \emph{Well-Being: Foundations of
  Hedonic Psychology}, D.~Kahneman, E.~Diener, and N.~Schwarz, Eds.\hskip 1em
  plus 0.5em minus 0.4em\relax New York, NY, USA: Utilitas, 1999, pp. 3--25.

\bibitem{Graziotin2014IEEESW}
D.~Graziotin, X.~Wang, and P.~Abrahamsson, ``{Software Developers, Moods,
  Emotions, and Performance.}'' \emph{IEEE Software}, vol.~31, no.~4, pp.
  24--27, 2014.

\bibitem{Vecchio:2000ei}
R.~P. Vecchio, ``{Negative Emotion in the Workplace: Employee Jealousy and
  Envy},'' \emph{International Journal of Stress Management}, vol.~7, no.~3,
  pp. 161--179, 2000.

\bibitem{Graziotin:2017}
D.~Graziotin, F.~Fagerholm, X.~Wang, and P.~Abrahamsson, ``Unhappy developers:
  Bad for themselves, bad for process, and bad for software product,'' in
  \emph{Proc.\ 2017 IEEE/ACM 39th International Conference on Software
  Engineering Companion (ICSE-C)}.\hskip 1em plus 0.5em minus 0.4em\relax ACM,
  2017.

\bibitem{Haybron2001}
D.~M. Haybron, ``{Happiness and Pleasure},'' \emph{Philosophy and
  Phenomenological Research}, vol.~62, no.~3, pp. 501--528, May 2001.

\bibitem{Diener2010}
E.~Diener, D.~Wirtz, W.~Tov, C.~Kim-Prieto, D.-w. Choi, S.~Oishi, and
  R.~Biswas-Diener, ``{New Well-being Measures: Short Scales to Assess
  Flourishing and Positive and Negative Feelings},'' \emph{Social Indicators
  Research}, vol.~97, no.~2, pp. 143--156, May 2010.

\bibitem{Haybron2005}
D.~M. Haybron, ``{On Being Happy or Unhappy},'' \emph{Philosophy and
  Phenomenological Research}, vol.~71, no.~2, pp. 287--317, Sep. 2005.

\bibitem{Graziotin2015SSE}
D.~Graziotin, X.~Wang, and P.~Abrahamsson, ``{Understanding the affect of
  developers: theoretical background and guidelines for psychoempirical
  software engineering},'' in \emph{SSE 2015: Proceedings of the 7th
  International Workshop on Social Software Engineering}.\hskip 1em plus 0.5em
  minus 0.4em\relax ACM, Sep. 2015, pp. 25--32.

\bibitem{graziotin2015affect}
------, ``{The Affect of Software Developers: Common Misconceptions and
  Measurements},'' \emph{Proceedings of the Eighth International Workshop on
  Cooperative and Human Aspects of Software Engineering}, pp. 123--124, 2015.

\bibitem{ColomoPalacios:2011jq}
R.~Colomo-Palacios and C.~Casado-Lumbreras, ``{Using the Affect Grid to Measure
  Emotions in Software Requirements Engineering},'' \emph{Journal of Universal
  Computer Science}, vol.~17, no.~9, pp. 1281--1298, 2011.

\bibitem{Wrobel2013}
M.~R. Wrobel, ``{Emotions in the software development process},'' \emph{2013
  6th International Conference on Human System Interactions (HSI)}, pp.
  518--523, 2013.

\bibitem{Mantyla2016a}
M.~M{\"a}ntyl{\"a}, B.~Adams, G.~Destefanis, D.~Graziotin, and M.~Ortu,
  ``{Mining valence, arousal, and dominance: possibilities for detecting
  burnout and productivity?}'' in \emph{MSR '16: Proceedings of the 13th
  International Conference on Mining Software Repositories}, Brunel
  University.\hskip 1em plus 0.5em minus 0.4em\relax New York, New York, USA:
  ACM, May 2016, pp. 247--258.

\bibitem{Creswell2009}
J.~W. Creswell, \emph{{Research design: qualitative, quantitative, and mixed
  method approaches}}, 3rd~ed.\hskip 1em plus 0.5em minus 0.4em\relax Thousand
  Oaks, California: Sage Publications, 2009, vol. 2nd.

\bibitem{Gousios:2016hj}
G.~Gousios, M.-a. Storey, and A.~Bacchelli, ``{Work practices and challenges in
  pull-based development},'' in \emph{Proc.\ 38th International Conference on
  Software Engineering - ICSE '16}, 2016, pp. 285--296.

\bibitem{Doll:2015uv}
\BIBentryALTinterwordspacing
B.~Doll. (2015) {A closer look at Europe}. [Online]. Available:
  \url{https://github.com/blog/2023-a-closer-look-at-europe}
\BIBentrySTDinterwordspacing

\bibitem{Graziotin:2016ck}
D.~Graziotin, F.~Fagerholm, X.~Wang, and P.~Abrahamsson, ``{Online appendix:
  the happiness of software developers},'' \emph{figshare}, 2017,
  https://doi.org/10.6084/m9.figshare.c.3355707.

\bibitem{Corbin2008}
J.~M. Corbin and A.~L. Strauss, \emph{{Basics of Qualitative Research:
  Techniques and Procedures for Developing Grounded Theory}}, 3rd~ed.\hskip 1em
  plus 0.5em minus 0.4em\relax London: Sage Publications, 2008, vol. 2nd.

\bibitem{Stol:2016:GTS:2884781.2884833}
K.-J. Stol, P.~Ralph, and B.~Fitzgerald, ``Grounded theory in software
  engineering research: A critical review and guidelines,'' in
  \emph{Proceedings of the 38th International Conference on Software
  Engineering}, ser. ICSE '16.\hskip 1em plus 0.5em minus 0.4em\relax New York,
  NY, USA: ACM, 2016, pp. 120--131.

\bibitem{Curtis:1988:FSS:50087.50089}
B.~Curtis, H.~Krasner, and N.~Iscoe, ``A field study of the software design
  process for large systems,'' \emph{Communications of the ACM}, vol.~31,
  no.~11, pp. 1268--1287, Nov. 1988.

\bibitem{Mitchell1982}
{Mitchell, T R}, ``{Motivation: New Directions for Theory, Research, and
  Practice},'' vol.~7, no.~1, pp. 80--88, 1982.

\bibitem{Franca2014b}
C.~Fran{\c c}a, H.~Sharp, and F.~da~Silva, ``{Motivated software engineers are
  engaged and focused, while satisfied ones are happy},'' in \emph{Proceedings
  of the 8th ACM/IEEE International Symposium on Empirical Software Engineering
  and Measurement}.\hskip 1em plus 0.5em minus 0.4em\relax New York, New York,
  USA: ACM Press, 2014, pp. 1--8.

\bibitem{Miner2010}
A.~G. Miner and T.~M. Glomb, ``{State mood, task performance, and behavior at
  work: A within-persons approach},'' \emph{Organizational Behavior and Human
  Decision Processes}, vol. 112, no.~1, pp. 43--57, May 2010.

\bibitem{Csikszentmihalyi1997}
M.~Csikszentmihalyi, \emph{{Finding flow: The psychology of engagement with
  everyday life}}, 1st~ed., ser. Personnel.\hskip 1em plus 0.5em minus
  0.4em\relax New York, New York, USA: Basic Books, 1997, vol.~30.

\bibitem{Ortu:2016gz}
M.~Ortu, G.~Destefanis, S.~Counsell, S.~Swift, M.~Marchesi, and R.~Tonelli,
  ``{How diverse is your team? Investigating gender and nationality diversity
  in GitHub teams},'' \emph{Peerj Preprints}, no.~4, p. e2285v1, Jul. 2016.

\bibitem{Djamba:2002jo}
Y.~K. Djamba and W.~L. Neuman, ``{Social Research Methods: Qualitative and
  Quantitative Approaches},'' \emph{Teaching Sociology}, vol.~30, no.~3, p.
  380, Jul. 2002.

\bibitem{Glaser:2013ha}
J.~Gl{\"a}ser and G.~Laudel, ``Life with and without coding: Two methods for
  early-stage data analysis in qualitative research aiming at causal
  explanations,'' \emph{Forum: Qualitative Social Research}, 2013.

\bibitem{Terrell:2016dq}
J.~Terrell, A.~Kofink, J.~Middleton, C.~Rainear, E.~Murphy-Hill, C.~Parnin, and
  J.~Stallings, ``{Gender differences and bias in open source: Pull request
  acceptance of women versus men},'' \emph{Peerj Preprints}, no.~4, p. e1733v2,
  Jul. 2016.

\bibitem{Ford:2016}
D.~Ford, J.~Smith, P.~J. Guo, and C.~Parnin, ``Paradise unplugged: Identifying
  barriers for female participation on stack overflow,'' in \emph{In Proc.\
  24th ACM SIGSOFT International Symposium on Foundations of Software
  Engineering}.\hskip 1em plus 0.5em minus 0.4em\relax New York, NY, USA: ACM,
  2016, pp. 846--857.

\end{thebibliography}

\end{document}